\documentclass{iopjournal}

\usepackage{amsmath,amssymb,bm}
\usepackage{booktabs,tabularx,array,multirow}
\usepackage[numbers,sort&compress]{natbib}
\usepackage{microtype}
\usepackage{siunitx}
\usepackage{caption}
\usepackage{subcaption}
\usepackage{enumitem}
\usepackage{placeins}

\graphicspath{{./}{figures/}}

\newlength{\doublepanelheight}
\hypersetup{colorlinks=true,linkcolor=blue,citecolor=blue,urlcolor=blue}
\newcommand{\Msun}{M_{\odot}}
\newcommand{\kpar}{\kappa_{\parallel}}
\newcommand{\kperp}{\kappa_{\perp}}

\newcommand{\CPCjournalname}{Chinese Physics C}
\newcommand{\CPCshortauthors}{Zhang et al.}

\fancyhfoffset[L,R]{0pt}

\makeatletter
\renewcommand{\articletype}[1]{%
  \vspace*{-8mm}%
  \noindent{\Large\sffamily \CPCjournalname\par}%
  \vspace*{6mm}%
  \noindent{\scriptsize\sffamily\bfseries
  \MakeUppercase{#1}\hfill\normalfont\itshape Submitted to Chinese Physics C\par}%
  \vspace*{4mm}%
}
\makeatother

\begin{document}

\articletype{Regular Article}

\title{Nuclear equation-of-state effects on the two-dimensional post-outburst thermal evolution of magnetized neutron-star crusts$^{*}$}

\author{Wei-Feng Zhang$^{1,2}$, Zhi-Fu Gao$^{1,\dagger}$, Wen-Qi Ma$^{1,2}$, Na Wang$^{1}$ and Yan-ling Liu$^{1}$}

\affil{$^1$ State Key Laboratory of Radio Astronomy and Technology, Xinjiang Astronomical Observatory, Chinese Academy of Sciences, Urumqi, Xinjiang 830011, China}

\affil{$^2$ University of Chinese Academy of Sciences, Beijing 100049, China}

\begingroup
\renewcommand{\thefootnote}{\fnsymbol{footnote}}
\footnotetext[1]{Supported by the National Natural Science Foundation of
China (NSFC) project No. 12288102, the National Key Research and Development Program of China (2022YFC2205202), the Major Science and Technology Special Project of Xinjiang Uygur Autonomous Region (2022A03013-1), the NSFC (12041304, 12573052, 12573103, 12373114, 12003009), the Natural Science Foundation of Xinjiang Uygur Autonomous Region (2022D01A155), and the Tianshan Talents Program
(2023TSYCTD0013).}
\footnotetext[2]{E-mail: zhifugao@xao.ac.cn}
\endgroup


\begin{abstract}

We present a controlled two-dimensional study of nuclear-equation-of-state (EOS) effects on the post-outburst thermal relaxation of magnetized neutron-star crusts. Six EOS models are evolved at a fixed gravitational mass of $1.4\,M_{\odot}$ with EOS-specific TOV backgrounds, crust compositions, and transport inputs under identical magnetic-field and heating prescriptions. Our analysis combines self-consistent multi-EOS evolution with a BSk-family factorization that separates structural, microphysical, and interaction contributions, together with an accepted-step energy ledger and a numerical-sensitivity budget. The models show EOS-dependent changes in both the early response and the later redistribution of heat between the surface, crust, and inner boundary; light-curve crossings near $10^2$ days demonstrate that the EOS effect is not a simple luminosity rescaling. By 1000 days the cumulative surface-photon energy fraction differs by less than one percentage point, whereas the internal energy partition differs much more strongly. The updated sensitivity tests show that the late-time luminosity differences and the largest peak contrasts exceed the corresponding BSk24 numerical-sensitivity scales, while the smallest early peak contrast remains less securely resolved. The calculations are intended as a reproducible EOS-sensitivity benchmark rather than an observational fit.

\end{abstract}

\section{Introduction}

Neutron stars connect nuclear physics at and above nuclear saturation density with observable high-energy astrophysics. Classic outer-crust and nucleonic-matter calculations, together with modern reviews, establish how the dense-matter equation of state (EOS) controls the mass--radius relation, composition, and crust--core structure \citep{Baym1971,Akmal1998,Lattimer2001,Haensel2007,Li2024}. Their thermal evolution additionally depends on heat capacity, neutrino emission, superfluidity, and transport throughout the core, crust, and heat-blanketing envelope \citep{Yakovlev2004,Page2004,Potekhin2015}. Recent unified EOS calculations have also demonstrated that EOS choice propagates into isolated-neutron-star cooling and crustal cooling in X-ray transients \citep{Tu2025}. In strongly magnetized objects, electron conduction becomes highly
anisotropic and produces nonuniform surface emission; the relevant envelope and crustal-transport foundations have been
developed and subsequently reviewed in
Refs.~\citep{Greenstein1983,Gudmundsson1983,Potekhin1997,
Potekhin1999,Potekhin2001,Geppert2004,Perna2001,Wang2025AN,Pons2026}.

Magnetar outbursts provide a natural setting for transient crustal heating. Their triggering may involve elastic or plastic crustal failure driven by the accumulation of magnetic stress \citep{Kojima2024}. Energy deposited near the surface or within the solid crust is redistributed by anisotropic conduction, photon emission, neutrino losses, and transport toward the core. Magnetic-field decay and magnetospheric return currents provide physically motivated heating channels \citep{Goldreich1992,Cumming2004,Pons2007,Beloborodov2016}, while magneto-thermal and source-oriented calculations demonstrate sensitivity to field geometry, heating depth, envelope composition, and crustal microphysics \citep{Pons2009,Vigano2013,Igoshev2021NatAs,Wang2021AN,Wang2024AN,Dehman2023MATINS,Ascenzi2024MATINS,DeGrandis2025}. However, the role of the nuclear EOS in two-dimensional post-outburst thermal evolution remains comparatively unexplored. Existing comparisons generally report the net response of self-consistent EOS models without isolating which differences are associated with stellar structure and which are associated with local crust microphysics. Establishing this separation is a prerequisite for assessing how EOS-dependent anisotropic thermal patterns may ultimately enter phase-resolved observational modelling; the present study itself does not compute pulse profiles or source-specific phase-resolved spectra. Here we therefore ask a controlled question: with stellar mass, magnetic topology, and heating prescription held fixed, how does changing the EOS alter the post-outburst response through the combined effects of stellar structure and crust microphysics?

Stable neutron-star fields can contain poloidal, toroidal, and higher-order components \citep{Braithwaite2006,Lander2009,Ciolfi2009}, and multidimensional magneto-thermal calculations show that field evolution can redistribute heat \citep{Dehman2023MATINS,Ascenzi2024MATINS,Pons2026}. To isolate EOS effects, we deliberately use the same static purely poloidal dipole in every run. Toroidal or multipolar structure and magnetic-field evolution are therefore outside the present controlled comparison and are discussed as model limitations.

We compare six EOS models using EOS-specific TOV backgrounds, crust compositions, and transport inputs at the same gravitational mass. The two-dimensional finite-volume calculation retains the projected symmetric conductivity tensor, including $K_{r\theta}$, applies a common fixed-fractional-depth heating prescription, and uses matched no-heating controls. A local-frame accepted-step ledger tracks the source, surface-photon energy, excess neutrino energy, inner-boundary energy exchange, and stored thermal energy with the same solver quadrature.

The paper is organized as follows. Section~\ref{sec:eos} describes the EOS-dependent structure and crust microphysics. Section~\ref{sec:model} presents the thermal model, heating prescription, and energy accounting. Section~\ref{sec:strategy} summarizes the numerical scheme together with the factorization and numerical-sensitivity diagnostics used in the analysis. Section~\ref{sec:results} presents the multi-EOS results, and Section~\ref{sec:outlook} gives the conclusions and limitations.

\section{EOS-dependent stellar structure and crust microphysics}
\label{sec:eos}

\subsection{Fixed-mass structure, composition, and common density support}

For each EOS, the hydrostatic background is obtained from the standard Tolman--Oppenheimer--Volkoff equations \citep{Lattimer2001,Haensel2007},
\begin{align}
\frac{dm}{dr} &= 4\pi r^2\rho, \\
\frac{dP}{dr} &= -\frac{G\left(\rho+P/c^2\right)\left(m+4\pi r^3P/c^2\right)}
{r^2\left(1-2Gm/rc^2\right)},
\label{eq:tov}
\end{align}
where $m(r)$ is the enclosed gravitational mass and $\rho$ is the mass-energy density divided by $c^2$. The central pressure is adjusted to obtain
\begin{equation}
M=1.4\,\Msun,
\label{eq:fixed_mass}
\end{equation}
thereby isolating EOS effects from stellar-mass differences. The resulting TOV solution determines the stellar radius $R$, crust--core transition radius $R_{\rm cc}$, crust thickness $\Delta R=R-R_{\rm cc}$, density profile, surface redshift
\begin{equation}
e^{\Phi_s}=\left(1-\frac{2GM}{Rc^2}\right)^{1/2},
\label{eq:redshift}
\end{equation}
and surface gravity
\begin{equation}
g_s=\frac{GM}{R^2e^{\Phi_s}}.
\label{eq:surface_gravity}
\end{equation}
Equations~(\ref{eq:redshift})--(\ref{eq:surface_gravity}) use the usual exterior Schwarzschild redshift and local surface-gravity expressions for a static spherical star \citep{Lattimer2001,Haensel2007}.
The complete TOV density profile is interpolated onto the thermal grid. The native TOV crust--core transition radius $R_{\rm cc}$ is retained as a structural quantity. The computational inner boundary $R_{\rm in}$ is placed at the largest density jointly supported by the structure EOS, the microphysics EOS, and the transport table; it coincides with $R_{\rm cc}$ only when that full native crust density range is supported.

Accordingly, $R_{\rm cc}$ and $R_{\rm in}$ have distinct roles in this work. The former characterizes the native stellar structure, whereas the latter defines the inner edge of the actually evolved thermal domain. Unless otherwise stated, volume-integrated crustal quantities in the thermal calculation refer to $R_{\rm in}\le r\le R$. Because the common table support can truncate the native crust before $R_{\rm cc}$ is reached, differences in the computational-domain depth must be kept conceptually separate from differences in the native crust thickness reported in Table~\ref{tab:structure}. The resulting run-specific $R_{\rm in}$ values and evolved-domain thicknesses $R-R_{\rm in}$ are reported in Table~\ref{tab:structure}.

Here ``EOS dependence'' denotes the full response to replacing the EOS at fixed gravitational mass. It includes both the resulting TOV structure---radius, compactness, surface gravity, native crust thickness, and density--radius mapping---and the EOS-dependent crust composition and transport inputs; it is therefore not reducible to a single soft--stiff ordering.

We consider BSk22, BSk24, BSk25, BSk26, SLy4mu, and CMF6mu. The four Brussels--Montreal BSk models are unified generalized-Skyrme energy-density-functional EOSs with outer crust, inner crust, and homogeneous core constructed within the same functional family \citep{Goriely2013BSk,Pearson2018,Pearson2019}. RG(SLy4)mu is an SLy4-based unified cold-star EOS using the Gulminelli--Raduta treatment of sub-saturation clustered matter \citep{Gulminelli2015,ComposeRGSLy42025}; the classic unified SLy4 construction provides the historical reference for this EOS family \citep{Douchin2001}. DS(CMF)-6mu uses the published CompOSE hybrid construction with a chiral mean-field CMF6 core and the RG(SkM) crust used in the present input set \citep{Dexheimer2008,Gulminelli2015,DanielewiczLee2009,ComposeCMF6Crust2025}. No BBP/BPS crust is used in the present input set. Because the models differ in calibration, composition, and crust construction as well as in their underlying theoretical frameworks, the comparisons below are model-to-model EOS comparisons rather than a pure SHF--RMF contrast. For clarity, the short labels SLy4mu and CMF6mu are used throughout the figures and discussion for the corresponding RG(SLy4)mu and DS(CMF)-6mu input constructions, respectively.

\begin{table*}[t]
\centering
\caption{EOS-dependent stellar structures and thermal-domain boundaries at $1.4\,\Msun$. The TOV quantities are derived from the cited EOS tables: BSk22, BSk24, BSk25, and BSk26 \citep{Goriely2013BSk,Pearson2018,Pearson2019}, SLy4mu \citep{Gulminelli2015,ComposeRGSLy42025,Douchin2001}, and CMF6mu \citep{Dexheimer2008,DanielewiczLee2009,ComposeCMF6Crust2025}. Here $R_{\rm cc}$ is the native crust--core transition radius, $R_{\rm in}$ is the actual computational inner-boundary radius, $R-R_{\rm in}$ is the thickness of the evolved thermal domain, $g_{14}=g_s/(10^{14}\ {\rm cm\,s^{-2}})$, and $e^{\Phi_s}$ is the surface redshift factor. Baryonic mass is not used as a control parameter.}
\label{tab:structure}
\scriptsize
\resizebox{\textwidth}{!}{%
\begin{tabular}{lcccccccc}
\toprule
EOS & $M/\Msun$ & $R_{\rm NS}$ (km) & $R_{\rm cc}$ (km) & $\Delta R$ (km) & $R_{\rm in}$ (km) & $R-R_{\rm in}$ (km) & $g_{14}$ & $e^{\Phi_s}$\\
\midrule
BSk22  & 1.400 & 13.047 & 11.953 & 1.094 & 11.9645 & 1.0827 & 1.321 & 0.8265\\
BSk24  & 1.400 & 12.587 & 11.552 & 1.035 & 11.5521 & 1.0350 & 1.431 & 0.8195\\
BSk25  & 1.400 & 12.385 & 11.397 & 0.987 & 11.3975 & 0.9871 & 1.484 & 0.8162\\
BSk26  & 1.400 & 11.773 & 10.843 & 0.930 & 10.8433 & 0.9301 & 1.664 & 0.8055\\
SLy4mu & 1.400 & 11.739 & 10.910 & 0.830 & 10.9206 & 0.8186 & 1.675 & 0.8049\\
CMF6mu & 1.400 & 13.342 & 12.604 & 0.738 & 12.6042 & 0.7376 & 1.257 & 0.8307\\
\bottomrule
\end{tabular}%
}
\end{table*}

The EOS tables provide the local composition. The cell mass number and crustal proton fraction are
\begin{equation}
A_{\rm cell}=\frac{A_{\rm bound}}{1-X_n},
\label{eq:acell}
\end{equation}
and
\begin{equation}
Y_p=(1-X_n)\frac{Z}{A_{\rm bound}},
\label{eq:yp}
\end{equation}
with the baseline crust-domain charge-neutrality condition
\begin{equation}
Y_\mu=0,\qquad Y_e=Y_p.
\label{eq:charge_baseline}
\end{equation}
Equations~(\ref{eq:acell})--(\ref{eq:charge_baseline}) are algebraic definitions based on the EOS-tabulated bound-cluster quantities, free-neutron fraction, and charge neutrality; they do not introduce an independent composition model. The numerical composition inputs are taken from the EOS tables cited above. Equation~(\ref{eq:charge_baseline}) applies only to the evolved crustal transport domain. Muons present in the high-density parent EOS are already encoded in the EOS used for the TOV background, whereas core lepton fractions are not evolved in the crust-only thermal calculation. The active crust transport uses $Y_\mu=0$ and $Y_e=Y_p$ as in the adopted crust-domain convention. Composition consistency is verified before constructing the transport coefficients.

\subsection{Transport coefficients and neutrino emission}

The EOS profile supplies the local composition and free-neutron density, while the EOS-tagged transport tables provide the anisotropic conductivities and heat capacities on a $(\rho,T,B)$ grid. Within the modeled crust, the dripped-neutron effective-mass ratio $m_n^*/m_n$ is evaluated by piecewise-linear interpolation of the fixed 42-point $k_{F,n}$-dependent table implemented in the MATINS microphysics routines \citep{Ascenzi2024MATINS}. The tabulated interval spans $k_{F,n}=0$--$1.7307\ {\rm fm}^{-1}$ and $m_n^*/m_n=1.0000$--$0.7558$; outside this interval the endpoint values are used. The same effective-mass prescription is applied to all six EOS models, while the local $k_{F,n}$ follows from the EOS-dependent crust composition. The singlet-neutron critical temperature $T_{c,n}$ is likewise evaluated with the same implemented SFB pairing prescription in all runs.
Transport queries are restricted to the tabulated $(\rho,T,B)$ support. The conductivity treatment follows standard magnetized-electron transport calculations \citep{Potekhin1999,Potekhin2001,Potekhin2015,Wang2020Universe}.

The magnetic conductivity tensor is
\begin{equation}
\hat{\bm{\kappa}}=\kperp\bm{I}+(\kpar-\kperp)\bm{b}\bm{b}
+\kappa_H[\bm{b}]_{\times},
\label{eq:kappa_tensor}
\end{equation}

Here $\bm{I}$ is the identity tensor, $\bm{b}=\bm{B}/B$ is the unit vector along the magnetic field, $\kappa_{\parallel}$ and $\kappa_{\perp}$ are the thermal conductivities parallel and perpendicular to $\bm{B}$, respectively, and $\kappa_H$ is the Hall (antisymmetric) conductivity coefficient. The projected coefficients $K_{rr}$, $K_{\theta\theta}$, and $K_{r\theta}$ below are the symmetric conductivity components in the meridional plane.

For an axisymmetric poloidal field, the symmetric components entering the $(r,\theta)$ thermal equation are
\begin{align}
K_{rr} &= \kperp+(\kpar-\kperp)b_r^2, \\
K_{\theta\theta} &= \kperp+(\kpar-\kperp)b_\theta^2, \\
K_{r\theta} &= (\kpar-\kperp)b_rb_\theta.
\label{eq:projected_kappa}
\end{align}
For the adopted axisymmetric purely poloidal field, $B_\phi=0$ and the temperature gradient lies in the meridional $(r,\theta)$ plane. The Hall heat flux generated by the antisymmetric term is then purely azimuthal and has zero divergence under axisymmetry. It therefore does not enter the present two-dimensional thermal equation, whereas the symmetric mixed component $K_{r\theta}$ is retained explicitly.

The total heat capacity includes ion, electron, and dripped-neutron contributions, with neutron superfluid suppression and pair-breaking/formation treated consistently with standard cooling models \citep{Yakovlev2004,Page2004}.

The crustal neutrino emissivity is
\begin{equation}
Q_\nu=Q_{nn}+Q_{eA}+Q_{\rm pl}+Q_{\rm syn}
+Q_{e^+e^-}+Q_{\rm PBF},
\label{eq:qnu_channels}
\end{equation}
including neutron--neutron and electron--nucleus bremsstrahlung, plasmon decay, synchrotron emission, pair annihilation, and singlet-neutron Cooper-pair breaking and formation \citep{Yakovlev2004,Potekhin2015}. Each $Q_i$ is a local neutrino energy-loss rate per unit volume and $Q_\nu$ is their sum. The PBF emissivity is active below $T_{c,n}(\rho)$, is largest near the pairing transition, and is strongly suppressed for $T\ll T_c$ \citep{Shternin2021}. Thus $T<T_c$ does not imply a large PBF contribution throughout the paired phase. Individual neutrino channels are retained in the run diagnostics.

\section{Two-dimensional thermal model and energy accounting}
\label{sec:model}

\subsection{Heat equation, magnetic geometry, and boundary conditions}

The local thermal-balance and Fourier-conduction equations follow the standard neutron-star cooling formulation \citep{Yakovlev2004,Potekhin2015,Pons2026}; in the present solver they are used in the local Newtonian-form transport approximation described below.
\begin{equation}
C_V\frac{\partial T}{\partial t}
=-\nabla\cdot\bm{F}+Q_{\rm heat}-Q_\nu,
\label{eq:heat}
\end{equation}
with
\begin{equation}
\bm{F}=-\hat{\bm{\kappa}}\cdot\nabla T.
\label{eq:fourier}
\end{equation}

In Eqs.~(\ref{eq:heat})--(\ref{eq:fourier}), $T(r,\theta,t)$ denotes the local proper temperature measured in the fluid rest frame, $C_V$ is the local heat capacity per unit volume, $\bm{F}$ is the local conductive heat flux, and $Q_{\rm heat}$ and $Q_\nu$ are the local volumetric heating and neutrino-loss rates. The coordinates $r$ and $\theta$ are the areal radius and magnetic colatitude used by the spherical finite-volume grid. For comparison with the general-relativistic thermal-equilibrium condition, let $e^{\Phi(r)}$ denote the lapse/redshift factor of the TOV background and define
\begin{equation}
\widetilde{T}(r,\theta,t)\equiv e^{\Phi(r)}T(r,\theta,t),
\label{eq:redshifted_temperature}
\end{equation}
which is spatially constant in an isothermal static configuration in full general relativity (Tolman condition) \citep{Potekhin2015,Pons2026}. The present solver instead evolves local $T$ with the Newtonian-form divergence operator in Eq.~(\ref{eq:heat}) on an EOS-dependent TOV background. It is therefore a partially relativistic treatment rather than a fully covariant thermal-transport calculation.

The flux components are
\begin{align}
F_r &=-K_{rr}\frac{\partial T}{\partial r}
-K_{r\theta}\frac{1}{r}\frac{\partial T}{\partial\theta},\\
F_\theta &=-K_{r\theta}\frac{\partial T}{\partial r}
-K_{\theta\theta}\frac{1}{r}\frac{\partial T}{\partial\theta}.
\label{eq:flux_components}
\end{align}

The baseline field is a static, axisymmetric, purely poloidal dipole. The adopted field prescription supplies $B_r(r,\theta)$ and $B_\theta(r,\theta)$ throughout the thermal domain, with $B_\phi=0$, and is normalized to the polar surface strength
\begin{equation}
B_p=5\times10^{14}\ {\rm G}.
\label{eq:Bpole}
\end{equation}

The thermal diffusion operator is evaluated in local spherical coordinates on the TOV-derived density mapping. General relativity enters through the EOS-specific TOV structure, the density--radius mapping, the surface gravity used in the envelope relation, and the redshift applied to the reported photon luminosities. Interior lapse factors, radial metric factors, and general-relativistic proper-volume factors are not included in the diffusion operator or in the local-frame energy quadrature. This approximation is stated explicitly because compactness is EOS dependent and therefore constitutes a possible systematic in a precision comparison among EOS models; its implications are revisited in Sec.~\ref{sec:outlook}.

The code uses the core-threading dipole prescription
\begin{align}
B_r(r,\theta)&=B_p\left(\frac{R}{r}\right)^3\cos\theta,\\
B_\theta(r,\theta)&=\frac{B_p}{2}\left(\frac{R}{r}\right)^3\sin\theta,\\
B_\phi(r,\theta)&=0.
\label{eq:dipole_field}
\end{align}
Equation~(\ref{eq:dipole_field}) is an imposed analytic dipole prescription, not a self-consistent magnetic-equilibrium solution. The same field geometry and normalization are used for every EOS model; only the EOS-dependent stellar structure changes the physical radial mapping of the thermal grid. Magnetic-field evolution, a toroidal component, and higher multipoles are excluded so that the calculation isolates the EOS dependence of the thermal response to a prescribed outburst.

At the inner boundary,
\begin{equation}
T(R_{\rm in},\theta,t)=T_{\rm core}=10^8\ {\rm K},
\label{eq:inner_bc}
\end{equation}
where $R_{\rm in}$ is the computational inner boundary. The imposed $T_{\rm core}=10^8$~K is a fixed \emph{local proper} boundary temperature; the core itself is not thermally evolved. Because $\Phi(R_{\rm in})$ depends on the EOS, fixing the same local $T_{\rm core}$ does not in general impose the same redshifted boundary temperature $\widetilde{T}_{\rm core}=e^{\Phi(R_{\rm in})}T_{\rm core}$ for all models. The resulting baseline should therefore be interpreted as a fixed-local-temperature boundary experiment, not as a fixed-$\widetilde{T}_{\rm core}$ experiment. At the outer boundary, the conductive flux is matched to an iron magnetic-envelope relation \citep{Gudmundsson1983,Potekhin1997,Potekhin2001,Greenstein1983}. The redshifted bolometric luminosity is \citep{Potekhin2015,Pons2026}
\begin{equation}
L_{\rm bol}^{\infty}
=e^{2\Phi_s}R^2\int F_s(\theta)\,d\Omega.
\label{eq:lbol_inf}
\end{equation}

Here $F_s(\theta)$ is the local outward surface photon flux supplied by the envelope boundary condition and $d\Omega$ is the solid-angle element. The factor $e^{2\Phi_s}$ converts the local surface luminosity to the luminosity measured by a distant observer within the adopted surface-redshift treatment.

\subsection{Heating prescription, control subtraction, and energy ledger}

We adopt a phenomenological two-component heating prescription. The scale $E_{\rm heat}=10^{41}$ erg and associated time/angle parameters are fiducial inputs chosen for this controlled benchmark, not values fitted to a particular source; they are representative of magnetar-heating and outburst calculations \citep{Beloborodov2016,DeGrandis2025}.
\begin{equation}
E_{\rm heat}=10^{41}\ {\rm erg},
\label{eq:Eheat}
\end{equation}
with surface fraction $f_s=0.4$. The temporal powers are
\begin{align}
P_s(t)&=\frac{f_sE_{\rm heat}}{\tau_s}e^{-t/\tau_s},\\
P_c(t)&=\frac{(1-f_s)E_{\rm heat}}{\tau_c}e^{-t/\tau_c},
\label{eq:heating_time}
\end{align}
where $\tau_s=15$ days and $\tau_c=120$ days. The angular profiles are Gaussian functions centered on the northern magnetic pole, with widths $\sigma_{\theta,s}=10^\circ$ and $\sigma_{\theta,c}=15^\circ$. The crustal radial profile is specified by the normalized depth
\begin{equation}
\xi=\frac{R-r}{R-R_{\rm in}},\qquad
w_\xi(r)=\exp\left[-\frac{(\xi-\xi_h)^2}{2\sigma_\xi^2}\right],
\label{eq:xi_heating}
\end{equation}
with $\xi_h=0.55$ and $\sigma_\xi=0.10$. This prescription places the heating at the same fractional depth in every EOS model and is the only radial prescription used in the results presented here.

Fixing $\xi_h$ controls the \emph{geometric fractional depth} of energy deposition, but it does not impose identical local density, pressure, composition, column depth, heat capacity, conductivity, pairing scale, or neutrino emissivity across EOS models. Those quantities remain EOS dependent at the heating location and are part of the response measured by the present fixed-$\xi$ experiment. Consequently, the baseline comparison should not be interpreted as a fixed-density or fixed-column-depth heating experiment.

Let $W_s(\theta)$ denote the surface angular weight and $W_c(r,\theta)=w_\xi(r)w_{\theta,c}(\theta)$ the crustal weight. The source terms are normalized separately for every EOS-dependent domain:
\begin{align}
F_{{\rm heat},s}(\theta,t)
&=P_s(t)\frac{W_s(\theta)}
{\displaystyle\int_{A_s}W_s\,dA},\\
Q_{{\rm heat},c}(r,\theta,t)
&=P_c(t)\frac{W_c(r,\theta)}
{\displaystyle\int_{V_{\rm crust}}W_c\,dV}.
\label{eq:heating_normalization}
\end{align}
Consequently, $\int_{A_s}F_{{\rm heat},s}\,dA=P_s(t)$ and $\int_{V_{\rm crust}}Q_{{\rm heat},c}\,dV=P_c(t)$. The implementation uses the same surface-ring areas and interior control-volume weights as the finite-volume solver, so EOS-dependent mesh geometry does not change the prescribed total source power.

The normalized coordinate $\xi=(R-r)/(R-R_{\rm in})$ places the source at the same fractional depth of the computational crust in every model and therefore avoids a trivial shift of the heating layer caused solely by different crustal thicknesses. The total injected energy is identical across the EOS models. Because the normalization uses the EOS-specific spherical control-volume weights of the discrete thermal operator, however, the local volumetric heating rate need not be identical: it depends on the weighted volume occupied by the prescribed source profile. These weights are local-frame discretization weights, not general-relativistic proper-volume elements.

Accordingly, the statement that all EOS models receive the same $E_{\rm heat}$ refers to the same prescribed \emph{local-frame source budget under the solver quadrature}. It does not assert equality of a redshifted injected energy measured at infinity. This distinction is immaterial for the internal bookkeeping performed by the present solver but is relevant when interpreting the calculation as a fully relativistic outburst-energy comparison.

Before the transient calculation, each EOS model is relaxed with $Q_{\rm heat}=0$ to a no-heating quasi-steady magnetized background using the same boundary conditions. The value $5\times10^7$~K is only the numerical seed for this relaxation. Heated and control runs then start from the same relaxed field $T_{\rm bg}(r,\theta)$, and the matched control removes residual baseline evolution and numerical drift. Excess quantities are defined by
\begin{equation}
X_{\rm ex}(t)=X_{\rm heated}(t)-X_{\rm control}(t),
\label{eq:control_subtraction}
\end{equation}
and this subtraction is applied consistently to photon luminosity, neutrino luminosity, and thermal energy. The excess crustal thermal energy is
\begin{equation}
E_{\rm th,ex}(t)
=\int_{V_{\rm crust}}
\left[u(T_{\rm heated})-u(T_{\rm control})\right]\,dV,
\label{eq:Eth_excess}
\end{equation}
using the same discrete control-volume weights as the finite-volume solver. For reference, the analytic cumulative energy specified by the two source laws is
\begin{equation}
E_{\rm inj}^{\rm ana}(t)=f_sE_{\rm heat}\left(1-e^{-t/\tau_s}\right)
 +(1-f_s)E_{\rm heat}\left(1-e^{-t/\tau_c}\right).
\label{eq:Einj_analytic}
\end{equation}
The energy ledger follows by summing the discrete finite-volume balance implied by Eq.~(\ref{eq:heat}) over each accepted step. The ledger is evaluated in the local frame of the accepted finite-volume update. Heated and matched-control contributions are computed with identical solver quadrature and subtracted step by step. The cumulative injected energy used by the ledger is the accepted-step discrete sum $E_{\rm inj}^{\rm disc}(t_N)=\sum_{n\leq N}\delta E_{\rm src,ex}^{(n)}$, so source normalization and closure use the same spatial quadrature and time sampling as the solver. Denoting the local control-subtracted source energy, surface-photon energy, excess neutrino energy, and inner-boundary energy exchange over one accepted step by $\delta E_{\rm src,ex}$, $\delta E_{\rm surf,ex}$, $\delta E_{\nu,\rm ex}$, and $\delta E_{\rm core,ex}$, respectively, the discrete identity is
\begin{equation}
\Delta U_{\rm ex}
=\delta E_{\rm src,ex}
-\delta E_{\rm surf,ex}
-\delta E_{\nu,\rm ex}
-\delta E_{\rm core,ex}
+\delta E_{\rm cl,ex},
\label{eq:energy_ledger_step}
\end{equation}
where positive $\delta E_{\rm core,ex}$ denotes energy leaving the simulated crust through the computational inner boundary, and $\delta E_{\rm cl,ex}$ is the discrete energy-closure residual. Here $E_{\rm surf,ex}$ denotes the cumulative excess surface-photon energy emitted through the outer boundary and is the cumulative counterpart of the surface photon luminosity. $E_{\nu,\rm ex}$ denotes the cumulative excess neutrino energy lost from the simulated crust, $E_{\rm core,ex}$ denotes the cumulative net energy transferred through $R_{\rm in}$, and $\Delta U_{\rm ex}$ is the excess thermal energy retained in the simulated crust. The symbol $E_{\rm core,ex}$ is retained as the ledger variable name, although the corresponding flux is evaluated at the computational inner boundary $R_{\rm in}$, which need not coincide with the native crust--core transition. Summed over all accepted steps,
\begin{equation}
E_{\rm inj}^{\rm disc}(t)
=E_{\rm surf,ex}(t)+E_{\nu,\rm ex}(t)+E_{\rm core,ex}(t)
+\Delta U_{\rm ex}(t)-E_{\rm cl,ex}(t).
\label{eq:energy_ledger}
\end{equation}
Photon light curves are reported separately as redshifted quantities $L_{\gamma,\rm ex}^{\infty}$. Fig~\ref{fig:energy_ledger} shows the local-frame ledger normalized by $E_{\rm inj}^{\rm disc}(1000\,{\rm d})$.

Across all baseline EOS runs, the cumulative energy-closure residual at 1000 days is $0.014\%$--$0.021\%$ of $E_{\rm inj}^{\rm disc}$, providing a direct measure of the numerical closure of the accepted-step ledger.

\begin{table}[!ht]
\centering
\caption{Baseline parameters adopted in this work. These are controlled simulation inputs rather than source-fitted measurements; the magnetar-scale field and heating choices are representative of the modelling literature \citep{Beloborodov2016,DeGrandis2025}.}
\label{tab:parameters}
\small
\begin{tabular}{lc}
\toprule
Parameter & Baseline value\\
\midrule
Gravitational mass & $1.4\,\Msun$\\
Polar magnetic field & $5\times10^{14}$ G\\
Local proper core-boundary temperature & $10^8$ K\\
Background-relaxation seed & $5\times10^7$ K\\
Physical crust initial state & EOS-specific no-heating quasi-steady background\\
Asymptotic local-frame prescribed source energy & $10^{41}$ erg\\
Surface heating fraction & $0.40$\\
Surface/crust heating timescales & $15$ d / $120$ d\\
Surface/crust angular widths & $10^\circ$ / $15^\circ$\\
Fixed fractional heating depth & $\xi_h=0.55$, $\sigma_\xi=0.10$\\
Evolution interval & $0$--$1000$ d\\
Baseline grid & $180\times60$\\
Time integrator & backward Euler\\
Baseline time-step prescription & $\Delta t=$ 0.001, 0.01, 0.10, 0.50, 2, 10 d over successive intervals\\
Transient nonlinear acceptance tolerances & $10^{-7}$ (Picard), $2\times10^{-6}$ ($L_2$), $2\times10^{-5}$ ($L_\infty$)\\
Envelope & iron\\
\bottomrule
\end{tabular}
\end{table}

Having defined the heating and energy ledger, we summarize the common physical and numerical parameters of the baseline runs in Table~\ref{tab:parameters}.

\section{Numerical method}
\label{sec:strategy}

Equation~(\ref{eq:heat}) is discretized conservatively on a uniform $(r,\theta)$ grid with spherical cell volumes and face areas \citep{Press2007,Barentzen2012,McCorquodale2015,White2024}. Positive diagonal conductivities are reconstructed at faces with a logarithmic mean. The $K_{rr}$ and $K_{\theta\theta}$ terms are treated implicitly, while the mixed flux is evaluated from the current Picard iterate,
\begin{align}
F_r^{{\rm cross},(k)}
&=-K_{r\theta}^{(k)}\frac{1}{r}\frac{\partial T^{(k)}}{\partial\theta},\\
F_\theta^{{\rm cross},(k)}
&=-K_{r\theta}^{(k)}\frac{\partial T^{(k)}}{\partial r}.
\label{eq:picard_cross}
\end{align}
This retains the cross flux without a fully implicit nine-point stencil. Backward Euler is used for the baseline calculations; the implementation also supports BDF2 for temporal checks. Neutrino losses are treated semi-implicitly while preserving a positive diffusion operator, and each step is accepted after the nonlinear update and residual tolerances are satisfied. The production time grid uses $\Delta t=0.001$, $0.01$, $0.10$, $0.50$, $2$, and $10$ days over the intervals 0--0.01, 0.01--0.1, 0.1--3, 3--30, 30--300, and 300--1000 days, respectively. The baseline transient acceptance tolerances are $10^{-7}$ for the Picard update, $2\times10^{-6}$ for the normalized $L_2$ residual, and $2\times10^{-5}$ for the normalized $L_\infty$ residual.

The six self-consistent EOS runs reveal the net EOS effect but conflate structural and microphysical contributions. To separate these pathways, additional off-diagonal calculations within the BSk family are used as a diagnostic factorization of structure and crust microphysics. Taking BSk24 as the reference EOS, the four quantities
\begin{equation}
X_{00}=X(S_0,M_0),\qquad
X_{i0}=X(S_i,M_0),\qquad
X_{0i}=X(S_0,M_i),\qquad
X_{ii}=X(S_i,M_i),
\label{eq:factorization_cells}
\end{equation}
are evaluated for $i\in\{\mathrm{BSk22},\mathrm{BSk25},\mathrm{BSk26}\}$. Here $S$ denotes the EOS used for the TOV structure and $M$ the EOS-tagged local crust-microphysics input set, including the EOS-dependent composition and the resulting transport and neutrino coefficients. The auxiliary neutron effective-mass and SFB pairing prescriptions are held fixed across the factorization matrix, while their local arguments remain EOS dependent. All cells are recomputed on one matrix-wide common density support shared by the BSk factorization set. The following decomposition is an algebraic inclusion--exclusion identity for the four calculated cells, rather than an independently assumed physical law.
The corresponding finite-difference decomposition is
\begin{align}
\Delta X_{\rm str}&=X_{i0}-X_{00},\\
\Delta X_{\rm mic}&=X_{0i}-X_{00},\\
\Delta X_{\rm int}&=X_{ii}-X_{i0}-X_{0i}+X_{00},
\label{eq:factorization_terms}
\end{align}
so that the total self-consistent difference satisfies
\begin{equation}
X_{ii}-X_{00}=\Delta X_{\rm str}+\Delta X_{\rm mic}+\Delta X_{\rm int}.
\label{eq:factorization_sum}
\end{equation}
This decomposition is a BSk24-referenced diagnostic rather than a unique, reference-independent causal partition.

BSk24 is selected as the reference because it provides a well-tested baseline already used in magneto-thermal calculations and lies within the structural range spanned by the BSk models considered here. The decomposition remains reference dependent and must not be interpreted as a unique causal partition. We therefore base claims of structural or microphysical dominance on the resolved pattern across the BSk comparisons, rather than on any single bar.

For the BSk24 baseline we also assess numerical sensitivity using three spatial resolutions ($90\times30$, $180\times60$, and $360\times120$), a halved-time-step test (all baseline physical time steps multiplied by 0.5), BDF2 time integration, and a tighter nonlinear test (all three transient acceptance tolerances reduced by a factor of 10). For an observable $Y$, the fractional deviation of a test run $q$ from the $180\times60$ backward-Euler baseline is defined by
\begin{equation}
\sigma_q(Y)=100\,\frac{|Y_q-Y_{\rm base}|}{|Y_{\rm base}|}.
\label{eq:sigma_q}
\end{equation}
The grid contribution is the larger of the coarse-grid and fine-grid deviations, while $\sigma_{\Delta t}$, $\sigma_{\rm int}$, and $\sigma_{\rm nl}$ are obtained from the halved-step, BDF2, and tighter-nonlinear runs, respectively. They are combined as
\begin{equation}
\sigma_{\rm num}(Y)=\left[\sigma_{\rm grid}^2(Y)+\sigma_{\Delta t}^2(Y)+\sigma_{\rm int}^2(Y)+\sigma_{\rm nl}^2(Y)\right]^{1/2}.
\label{eq:sigma_num}
\end{equation}
The quantity $\sigma_{\rm num}(Y)$ is used as a practical numerical-sensitivity indicator, not as a statistical confidence interval. The quadrature combination in Eq.~(\ref{eq:sigma_num}) is therefore only a compact summary of several controlled perturbations; it should not be interpreted as demonstrating statistical independence of grid, time-step, integrator, and nonlinear-solver effects. Throughout the discussion below, a contrast is described as numerically resolved only when its magnitude exceeds the corresponding $\sigma_{\rm num}$; contrasts at or below that scale are treated as unresolved or indicative. This is a numerical-resolution convention, not a statistical-significance criterion.

\section{Results and discussion}
\label{sec:results}

\subsection{EOS-dependent thermal response}

\begin{figure*}[!ht]
\centering
\begin{subfigure}[t]{0.485\textwidth}
  \centering
  \includegraphics[height=\doublepanelheight]{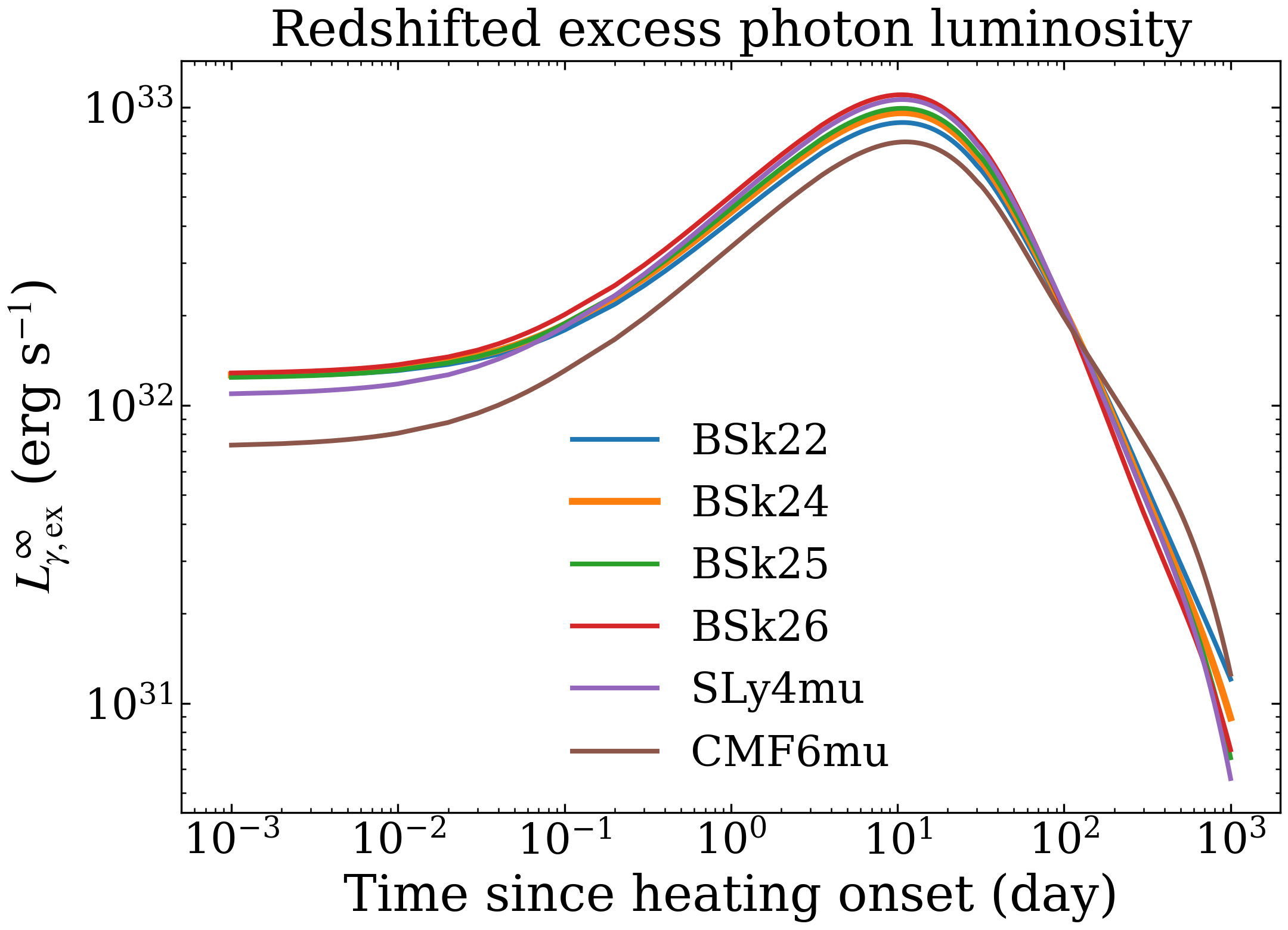}
  \caption{Redshifted excess photon luminosity.}
  \label{fig:luminosity_absolute}
\end{subfigure}
\hfill
\begin{subfigure}[t]{0.485\textwidth}
  \centering
  \includegraphics[height=\doublepanelheight]{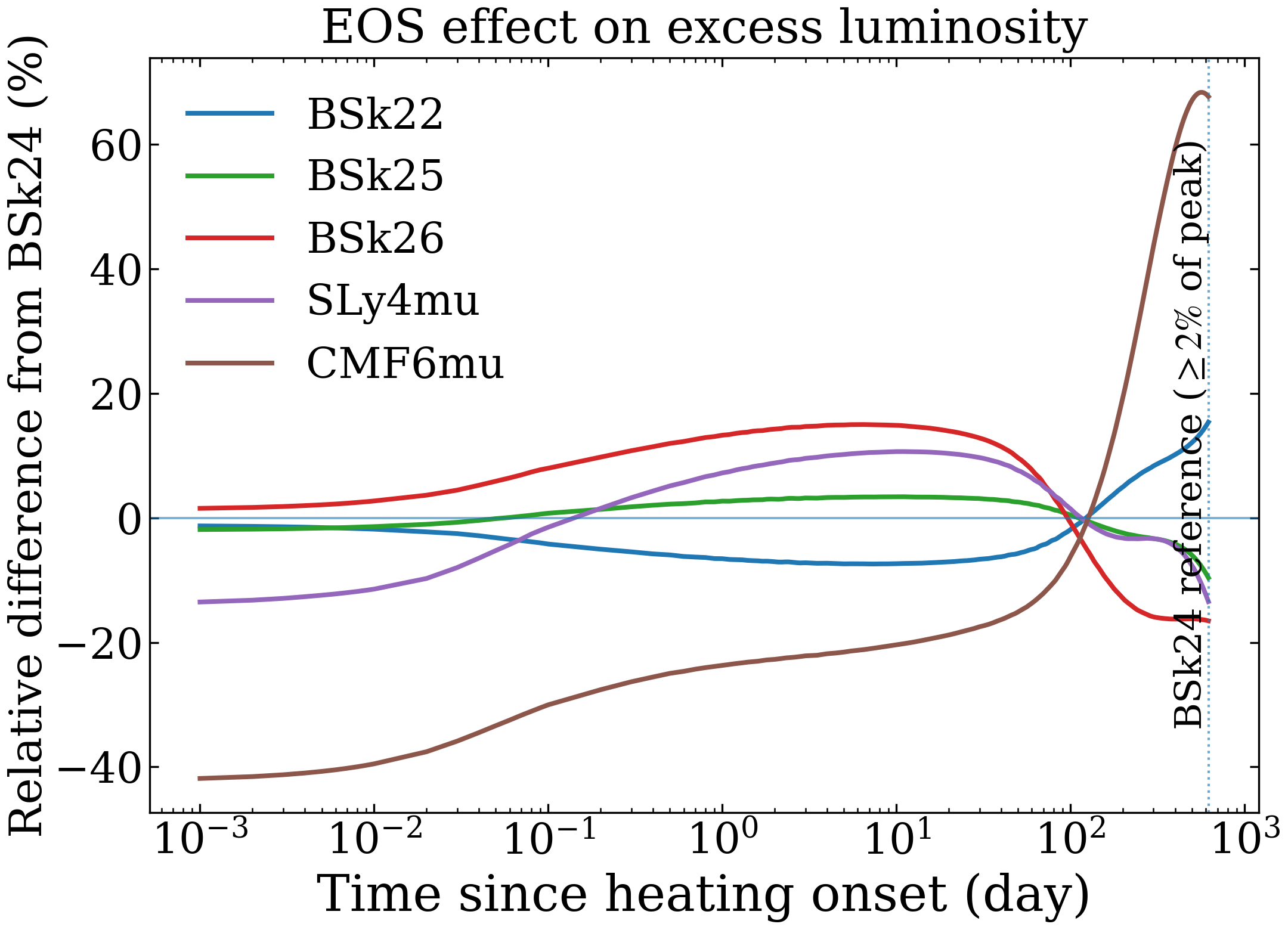}
  \caption{Difference relative to BSk24.}
  \label{fig:luminosity_relative}
\end{subfigure}
\caption{Simulation-derived EOS dependence of the redshifted excess photon luminosity. Broad maxima occur at $10.5$--$11.0$ days. Relative differences are shown only while the BSk24 excess exceeds $2\%$ of its peak, avoiding ratios dominated by a vanishing reference signal.}
\label{fig:luminosity}
\end{figure*}

\begin{figure}[!ht]
\centering
\includegraphics[width=0.6\columnwidth]{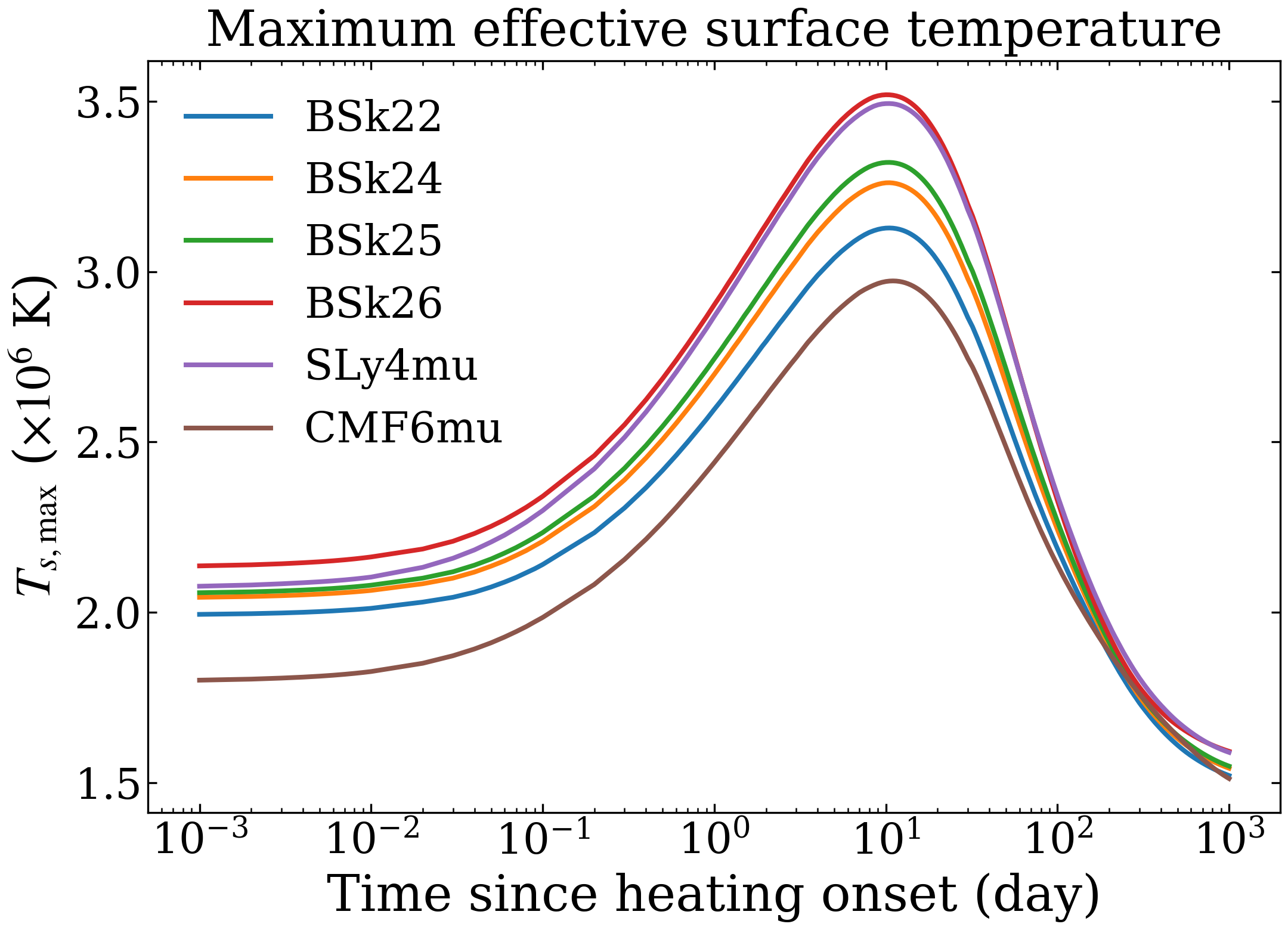}
\caption{Simulation-derived maximum local (proper-frame) effective surface temperature. The broad maximum occurs at $10.5$--$11$ days; BSk26 and SLy4mu give the strongest peak response and CMF6mu the weakest.}
\label{fig:surface_temperature_max}
\end{figure}

\begin{figure*}[!ht]
\centering
\begin{subfigure}[t]{0.485\textwidth}
  \centering
  \includegraphics[height=\doublepanelheight]{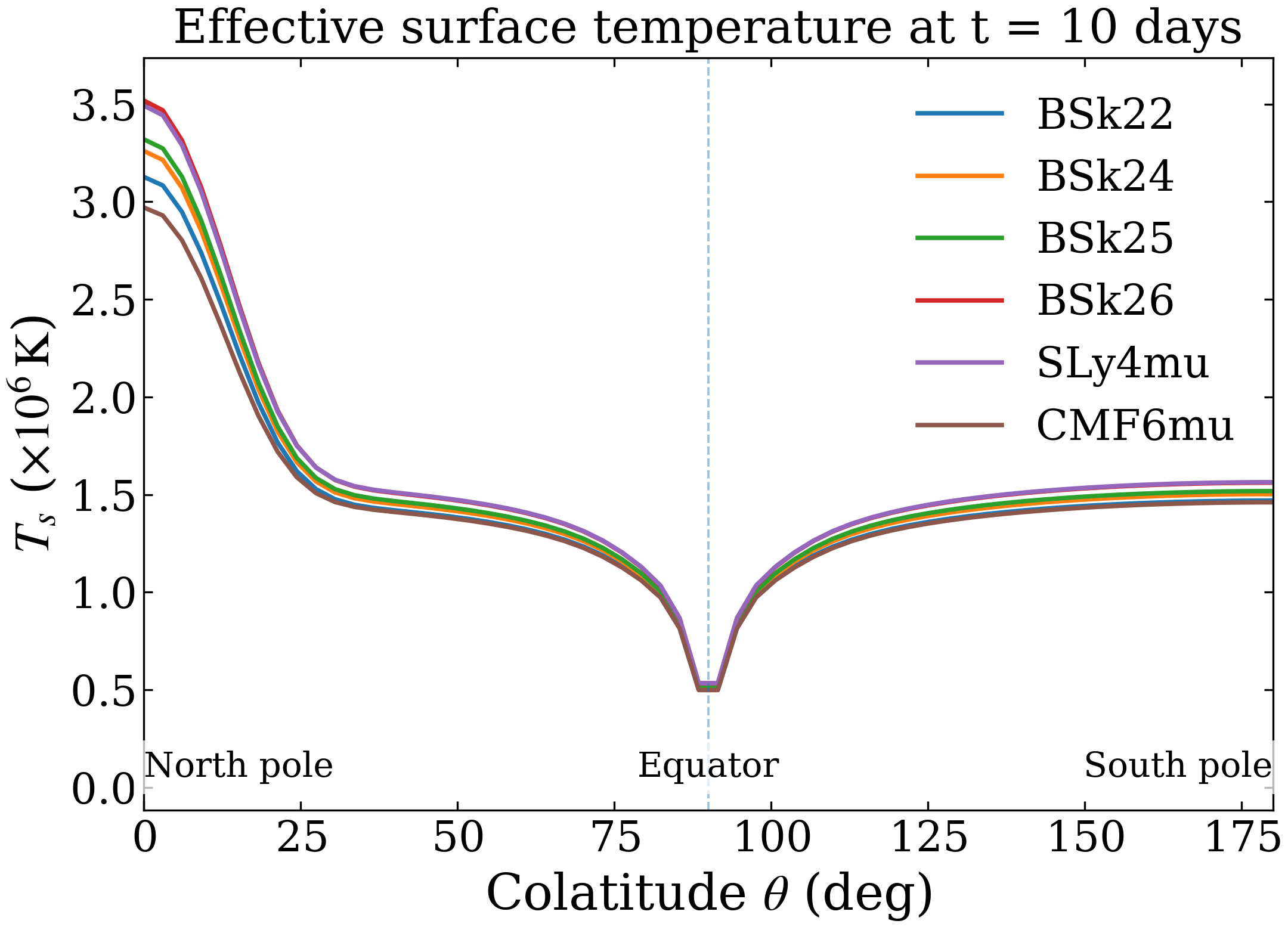}
  \caption{Surface profile at 10 days.}
  \label{fig:surface_profile_10d}
\end{subfigure}
\hfill
\begin{subfigure}[t]{0.485\textwidth}
  \centering
  \includegraphics[height=\doublepanelheight]{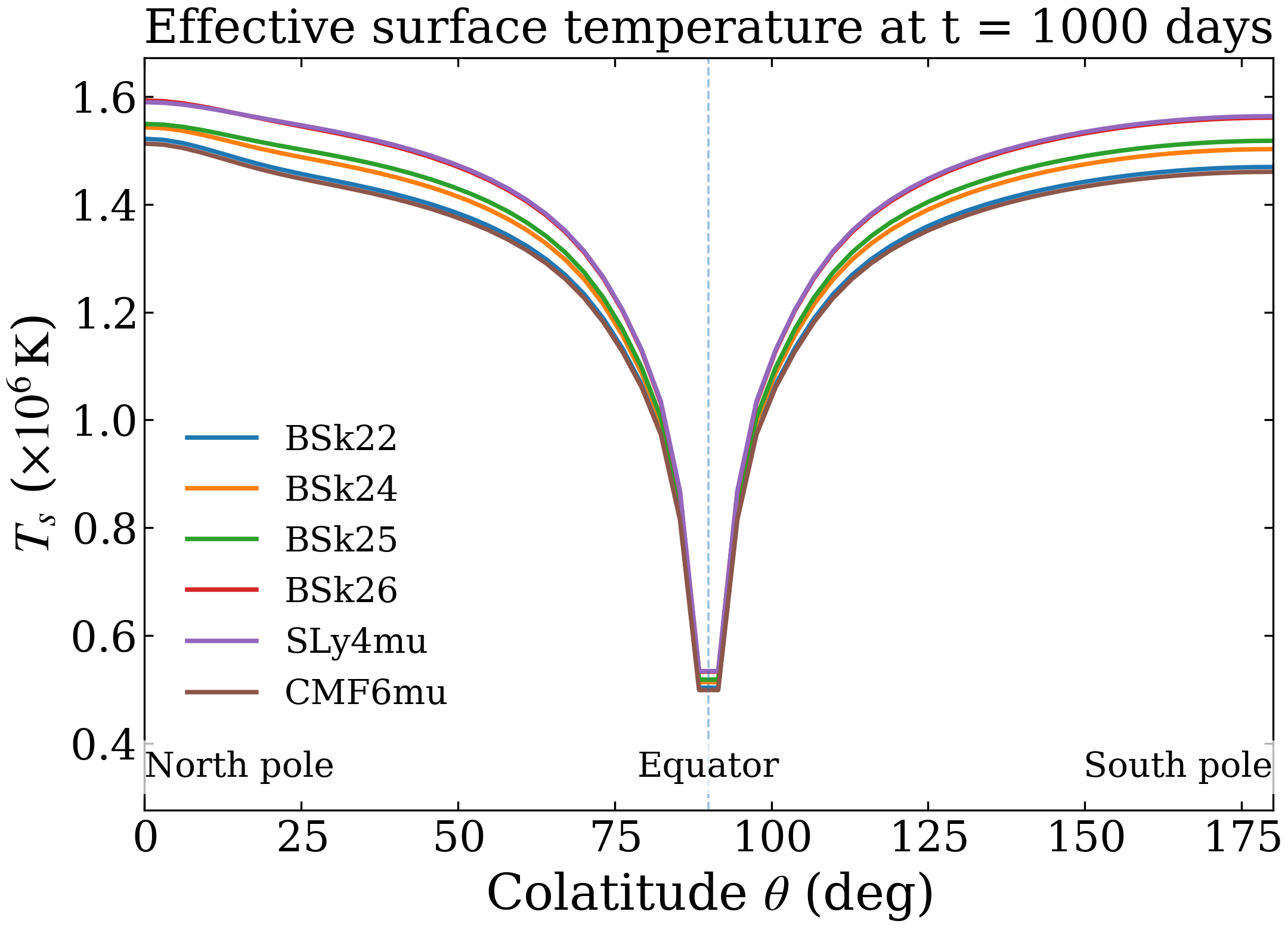}
  \caption{Surface profile at 1000 days.}
  \label{fig:surface_profile_1000d}
\end{subfigure}
\caption{Simulation-derived angular distributions of the local effective surface temperature at 10 and 1000 days. The northern hot cap largely relaxes, whereas the magnetically produced equatorial cold belt remains visible.}
\label{fig:surface_profiles}
\end{figure*}

All numerical values plotted in Figs.~\ref{fig:luminosity}--\ref{fig:numerical_sensitivity} are outputs of the simulations defined in Secs.~\ref{sec:eos}--\ref{sec:strategy} and Table~\ref{tab:parameters}; no observational data are plotted. The EOS input sources are given in Table~\ref{tab:structure}.

Table~\ref{tab:structure} summarizes the structural EOS dependence at the fixed gravitational mass of $1.4\,\Msun$. The stellar radii span $11.739$--$13.342$ km, the native crust--core transition radii span $10.843$--$12.604$ km, and the native crust thicknesses span $0.738$--$1.094$ km. BSk22 has the thickest native crust, whereas CMF6mu has the largest stellar radius but the thinnest native crust. BSk26 and SLy4mu are the most compact models and consequently have the largest surface gravities and the smallest surface redshift factors. These trends show that the thermal response cannot be inferred from the stellar radius alone: emitting area, gravitational redshift, native crust geometry, and the density--radius mapping vary simultaneously and can act in opposite directions.

Physically, these structural quantities enter different parts of the problem. The radius sets the emitting area, the compactness controls the surface-redshift factor and surface gravity entering the envelope mapping, and the EOS-dependent density--radius relation sets the physical distance over which heat diffuses between the deposition region, the surface, and the inner boundary. The present calculation does not isolate these effects one by one outside the BSk factorization; they should therefore be regarded as coupled structural influences rather than a single monotonic ``stiffness'' effect.

The EOS-dependent composition provides a second pathway. Differences in $Z$, $A_{\rm bound}$, $A_{\rm cell}$, $X_n$, $Y_e$, and the local neutron Fermi momentum propagate through the adopted common effective-mass and pairing prescriptions into the heat capacity, anisotropic conductivities, and channel-resolved neutrino emissivities. The self-consistent light curves therefore represent the combined response of global stellar structure and local crust microphysics.

At early times, the emergent peak is sensitive to how rapidly the deposited heat can be stored and conducted away from the heated layer; at later times, the competition between outward photon leakage, inward transport, and residual crustal storage controls the relaxation. This motivates interpreting the EOS-dependent curve crossings below primarily as changes in thermal-relaxation timescale, rather than only as changes in luminosity normalization.

Fig~\ref{fig:luminosity} compares the redshifted post-outburst excess luminosities. All six models reach broad maxima at $10.5$--$11.0$ days. The peak ordering is BSk26, SLy4mu, BSk25, BSk24, BSk22, and CMF6mu; relative to BSk24, BSk26 is $14.8\%$ higher and CMF6mu $20.2\%$ lower. The same ordering appears in peak $T_{\rm s,max}$ but is not preserved during the later relaxation.

The peak times and absolute multi-EOS peak values quoted as physical results in this paper are taken from Fig.~\ref{fig:luminosity}. The $180\times60$ BSk24 baseline shown in Fig.~\ref{fig:numerical_sensitivity}(a) is the same production BSk24 solution used in Fig.~\ref{fig:luminosity}; Fig.~\ref{fig:numerical_sensitivity} only adds numerical variants around this reference solution. It is therefore used solely to quantify resolution, time-integration, and solver sensitivity, and does not define a second physical peak time or replace the production multi-EOS light curves.

The curves cross near the $10^2$-day scale. Thereafter, BSk22 and CMF6mu decay more slowly relative to BSk24, while BSk25, BSk26, and SLy4mu fall below the reference. The relative excess luminosity is defined by
\begin{equation}
\delta_L^{(i)}(t)
=100\frac{L_{\gamma,\rm ex,i}^{\infty}(t)-L_{\gamma,\rm ex,BSk24}^{\infty}(t)}
{L_{\gamma,\rm ex,BSk24}^{\infty}(t)}.
\label{eq:relative_luminosity}
\end{equation}
The sign reversal relative to BSk24 shows that the EOS changes both the peak normalization and the relaxation timescale. A single multiplicative rescaling of the BSk24 light curve therefore cannot reproduce the full multi-EOS behavior. In physical terms, the crossing indicates that the model producing the larger early surface response need not retain the larger late surface flux: the relative importance of outward diffusion, inward transport, and thermal storage evolves with time and depends on both structure and microphysics.

The surface temperatures provide complementary information on the same EOS-dependent response.

The maximum surface-temperature histories in Fig.~\ref{fig:surface_temperature_max} follow the luminosity ordering around the peak, spanning $2.97\times10^6$ K for CMF6mu to $3.52\times10^6$ K for BSk26. By 1000 days the curves converge toward approximately $(1.5$--$1.6)\times10^6$ K, with a residual EOS ordering. The stronger convergence than in the excess luminosity reflects the additional dependence of bolometric emission on emitting area, angular redistribution, and gravitational redshift.

The angular profiles in Fig.~\ref{fig:surface_profiles} show that the angle-integrated light curve does not contain all of the thermal information. At 10 days, the one-sided heating produces a pronounced northern hot cap, whereas the magnetized envelope and anisotropic crustal conduction maintain a deep equatorial minimum near $0.5\times10^6$ K. By 1000 days, the original hemispheric asymmetry has largely relaxed and the two hemispheres become much closer, but the equatorial insulation persists.

The similar positions of the equatorial minima are consistent with the common magnetic geometry setting the broad angular pattern, while EOS changes mainly affect the amplitude and persistence of the polar excess. BSk26 and SLy4mu produce the highest early polar temperatures, whereas BSk22 and CMF6mu show slower late luminosity decay despite lower surface maxima. Local temperature, emitting area, and redshift therefore contribute differently to the bolometric light curve.

\subsection{Energy redistribution and neutrino emission}

\begin{figure*}[!ht]
\centering
\begin{subfigure}[t]{0.49\textwidth}
\centering
\includegraphics[width=\linewidth]{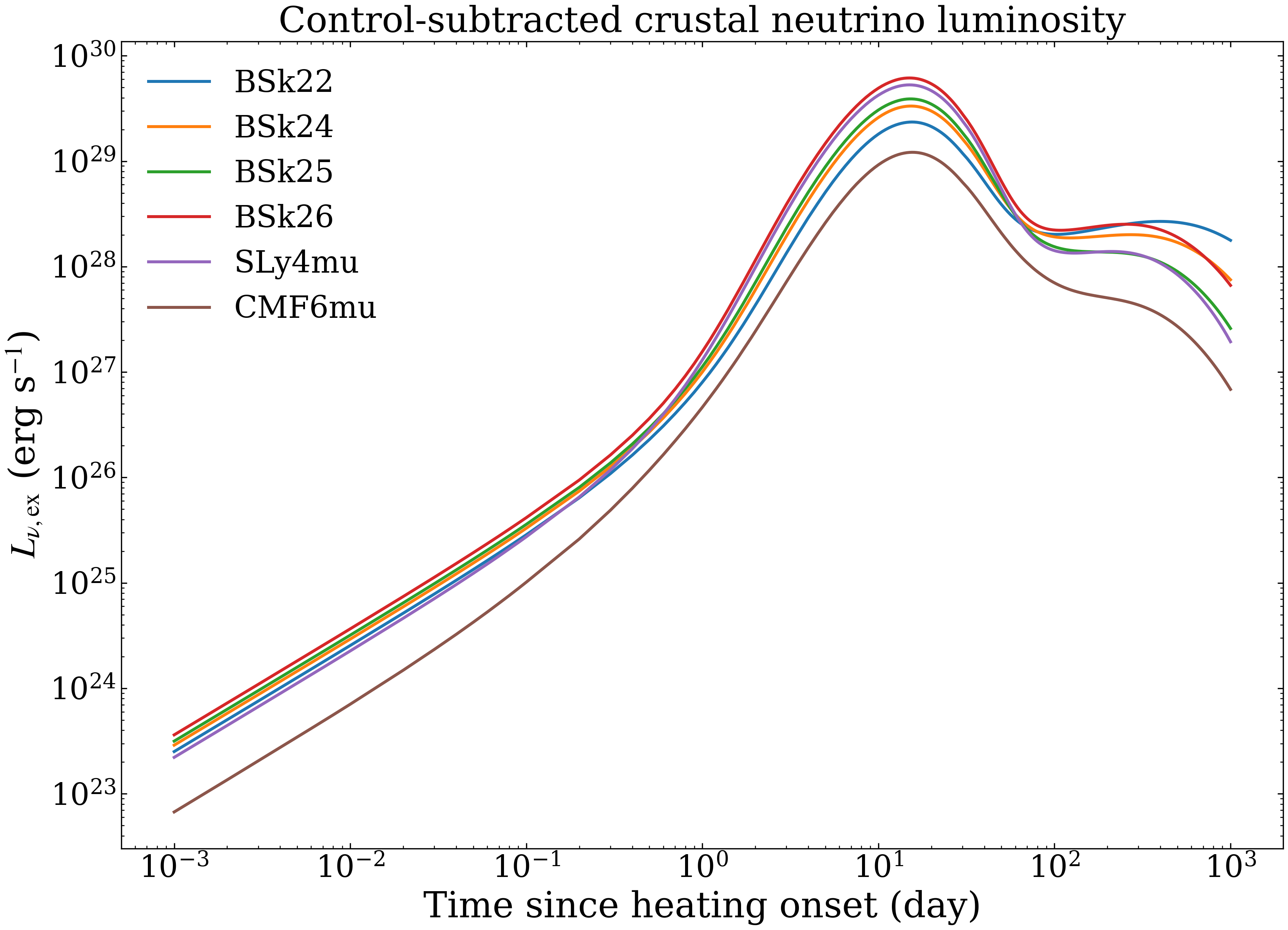}
\caption{Simulation-derived control-subtracted crustal neutrino luminosity, peaking at $15.0$--$15.5$ days.}
\label{fig:neutrino_excess}
\end{subfigure}\hfill
\begin{subfigure}[t]{0.5\textwidth}
\centering
\includegraphics[width=\linewidth]{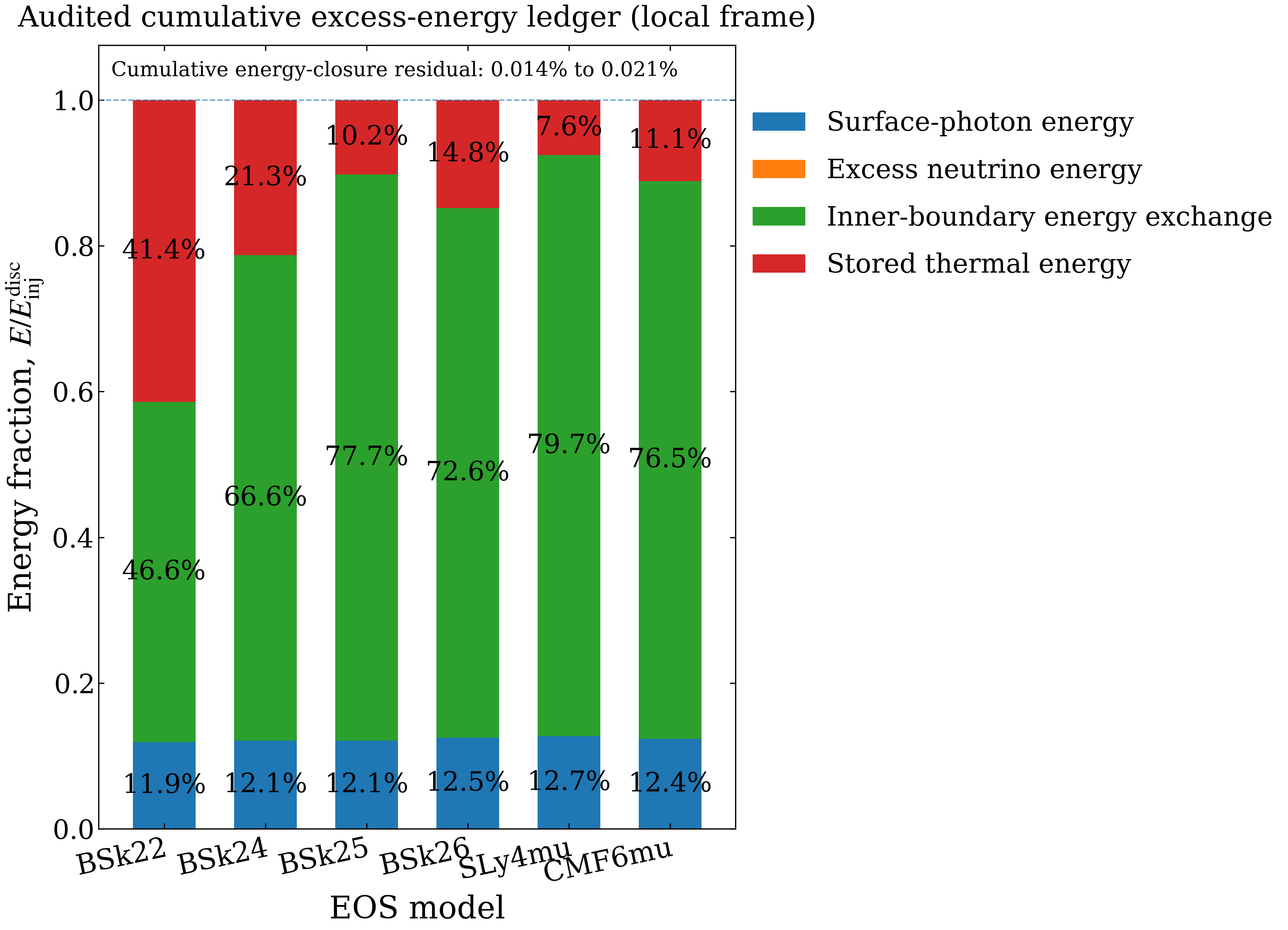}
\caption{Simulation-derived local-frame energy partition at 1000 days, normalized by $E_{\rm inj}^{\rm disc}$. ``Surface-photon energy'' is the cumulative surface-photon emission $E_{\rm surf,ex}$.}
\label{fig:energy_ledger}
\end{subfigure}

\caption{Neutrino response and discrete energy accounting from the present simulations. In panel (b), surface-photon energy denotes the cumulative excess surface-photon energy $E_{\rm surf,ex}$, positive inner-boundary energy exchange denotes energy leaving the simulated crust through $R_{\rm in}$, and the energy-closure residual denotes the cumulative $E_{\rm cl,ex}$; its magnitude is $0.014\%$--$0.021\%$.}
\label{fig:energy_diagnostics}
\end{figure*}

The control-subtracted crustal neutrino luminosity in Fig.~\ref{fig:neutrino_excess} peaks at $15.0$--$15.5$ days, with peak values spanning $1.21\times10^{29}$--$6.16\times10^{29}$ erg s$^{-1}$. At later times a broad shoulder develops between roughly $10^2$ and several $10^2$ days, with BSk22 declining most slowly. The total neutrino light curve combines the EOS dependence of temperature, composition, pairing, and the density regions sampled by the emissivity channels; no channel-by-channel attribution is attempted here.

Beyond photon luminosity, partitioning the deposited energy among surface-photon energy, excess neutrino energy, inner-boundary energy exchange, and stored thermal energy provides a more complete measure of the EOS-dependent thermal response.

By 1000 days, surface-photon energy accounts for $11.9\%$--$12.7\%$ of the cumulative accepted-step injected energy. Inner-boundary energy exchange spans $46.6\%$--$79.7\%$, while the crust retains $7.6\%$--$41.4\%$ as stored excess thermal energy. BSk22 has the smallest inner-boundary exchange and the largest stored fraction, whereas SLy4mu shows the opposite pattern. The cumulative excess neutrino energy is subdominant in the 1000-day ledger. Thus the surface-photon energy fraction varies by less than one percentage point, while the internal redistribution between inner-boundary exchange and crustal storage varies by tens of percentage points.

The surface-photon entry in the local-frame energy ledger is not itself a distant-observer quantity; its corresponding redshifted excess luminosity $L_{\gamma,\rm ex}^{\infty}$ is the quantity directly connected to electromagnetic observations. The neutrino term is retained here primarily as an internal cooling and energy-closure diagnostic rather than as a directly measurable observable for these sources. The comparatively narrow spread in the cumulative surface-photon energy fraction, together with the much broader spread in inner-boundary energy exchange and stored thermal energy, suggests that the strongest EOS sensitivity in this controlled experiment lies in how the crust redistributes energy internally. This interpretation remains conditional on the fixed local-temperature inner boundary and on the computational depth $R_{\rm in}$, both of which affect the inward-energy channel.

\FloatBarrier
\subsection{BSk structure--microphysics factorization}

\begin{figure*}[!ht]
\centering
\includegraphics[width=0.90\textwidth]{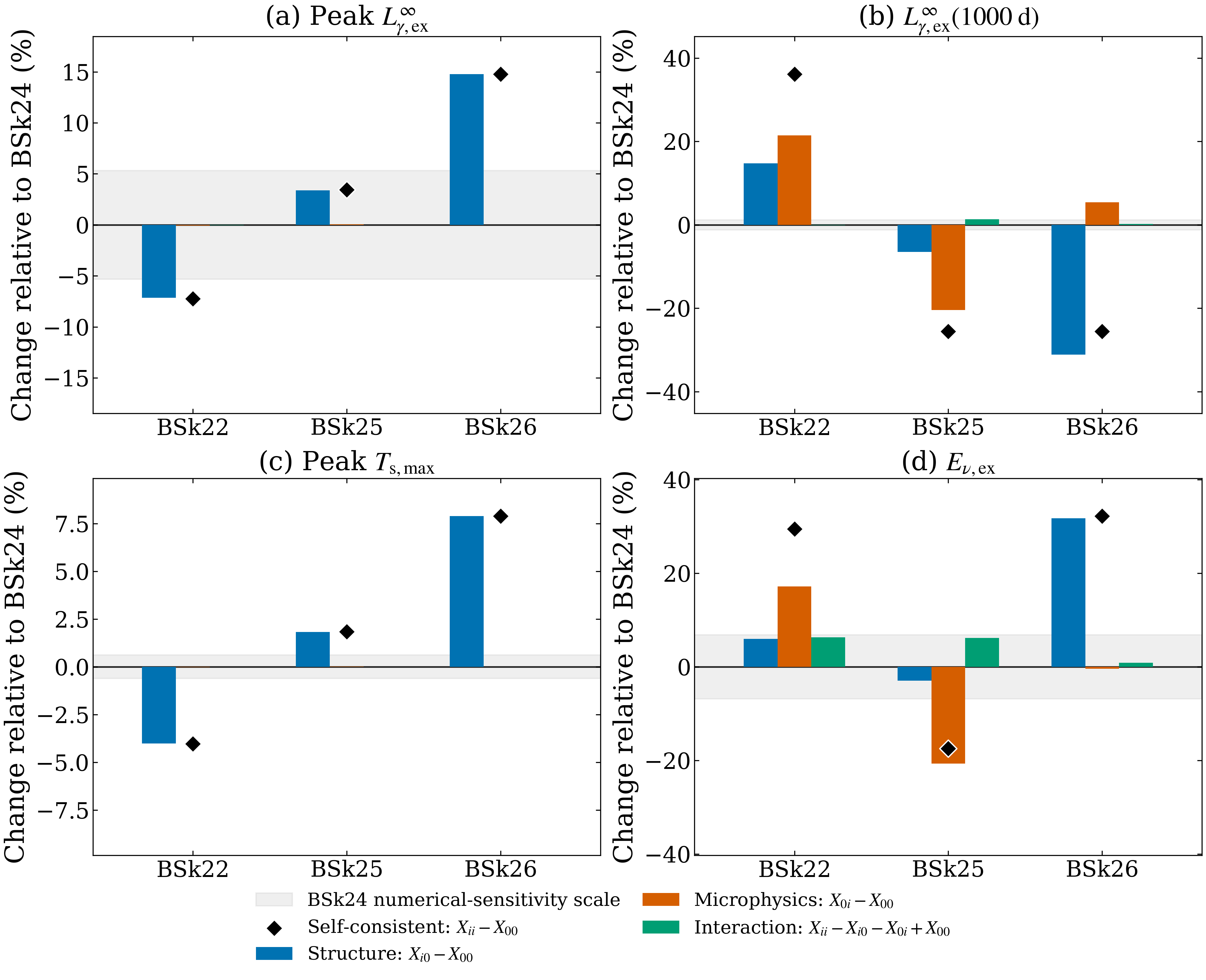}
\caption{Simulation-derived BSk24-referenced factorization into structure, microphysics, and interaction terms; black diamonds show the self-consistent differences $X_{ii}-X_{00}$. The four cells share a common density support. Panels show (a) peak redshifted excess photon luminosity, (b) $L_{\gamma,\rm ex}^{\infty}(1000\,{\rm d})$, (c) peak $T_{\rm s,max}$, and (d) cumulative excess neutrino energy. Gray bands show the corresponding BSk24 combined numerical-sensitivity scales from Fig.~\ref{fig:numerical_sensitivity}. The decomposition is reference dependent and diagnostic; differences lying within the gray band are treated as unresolved by the adopted reference-sensitivity criterion.}
\label{fig:factorization}
\end{figure*}

Fig~\ref{fig:factorization} separates the BSk24-referenced differences into structure, microphysics, and interaction contributions. For the peak excess photon luminosity and peak $T_{\rm s,max}$, the structural term sets almost the entire BSk variation, while the microphysical and interaction terms are small. Relative to the updated BSk24 sensitivity scales, the BSk26 and BSk22 peak-luminosity contrasts lie outside the gray band, whereas the smaller BSk25 peak-luminosity contrast remains within it. All three peak-$T_{\rm s,max}$ contrasts exceed the corresponding reference sensitivity scale. These comparisons support structural control of the early peak response within the BSk subset, while also showing that the smallest peak-luminosity contrast should not be over-interpreted.

At 1000 days the decomposition becomes strongly model dependent. For BSk22, both structure and microphysics increase the excess luminosity, with the microphysical contribution larger than the structural one. For BSk25, the late-time suppression is driven primarily by the negative microphysical contribution, with a smaller negative structural term and a weak positive interaction correction. For BSk26, the suppression is instead dominated by structure and is partly offset by a positive microphysical contribution. The interaction term remains secondary in all three 1000-day luminosity comparisons. Because the total late-time contrasts are much larger than the updated $1.20\%$ BSk24 sensitivity indicator for $L_{\gamma,\rm ex}^{\infty}(1000\,{\rm d})$, the EOS-dependent differences in relaxation are more securely resolved than the smallest early-peak difference.

The cumulative excess neutrino energy shows a different balance again. The microphysical term is the largest single contribution for BSk22 and BSk25, while the BSk26 enhancement is dominated by structure; interaction terms are non-zero but remain smaller than the leading contribution. The self-consistent differences for all three BSk alternatives exceed the updated $6.84\%$ BSk24 sensitivity scale for $E_{\nu,\rm ex}$. This result argues against describing the neutrino-energy response as generically structure dominated: within the present factorization, the relative importance of structure and microphysics depends on which BSk EOS is compared with BSk24.

\FloatBarrier
\subsection{Numerical sensitivity of the BSk24 baseline}

\begin{figure*}[!ht]
\centering
\includegraphics[width=0.94\textwidth]{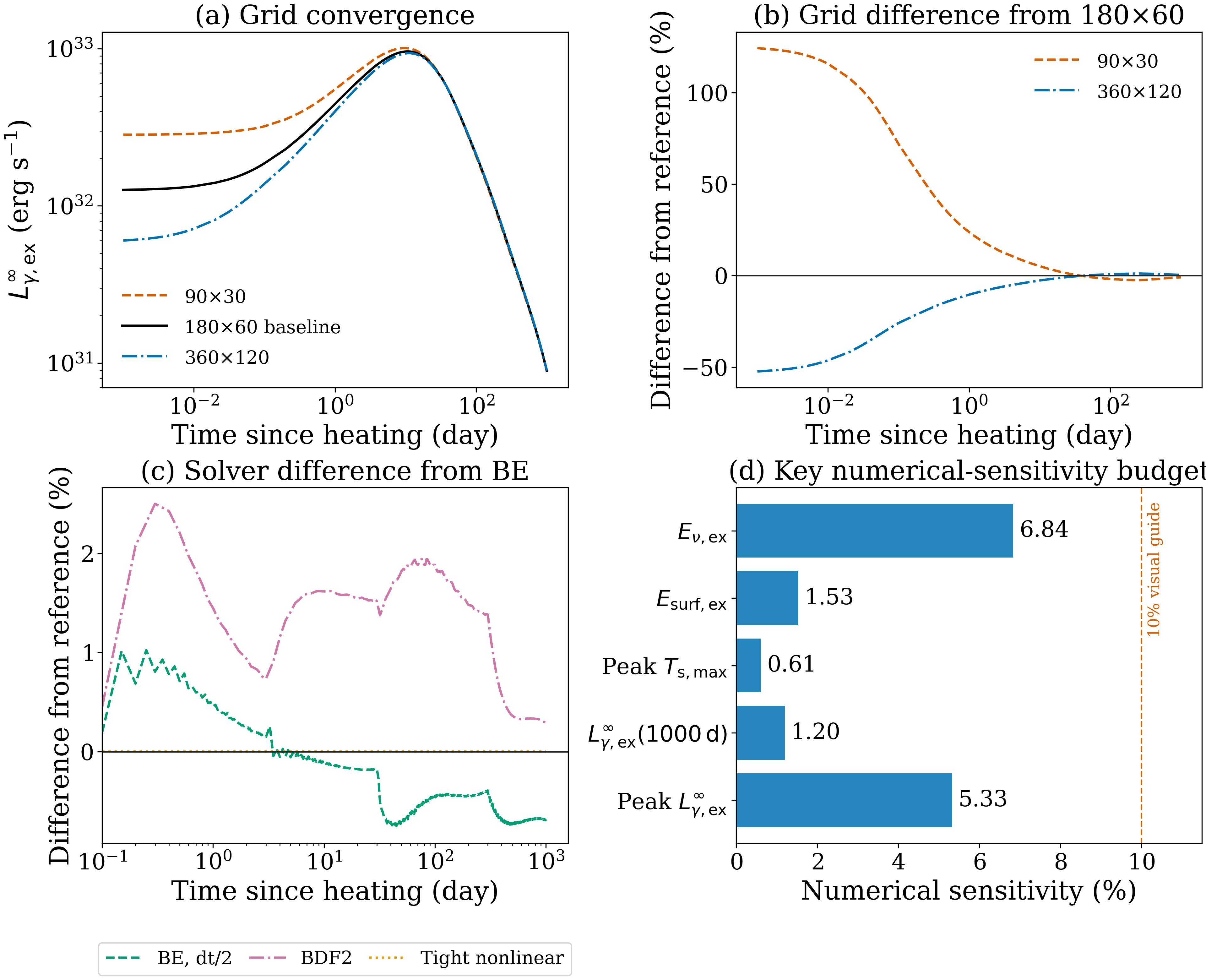}
\caption{Simulation-derived numerical-sensitivity tests for BSk24. Panels show (a) three grid resolutions, (b) grid differences relative to $180\times60$, (c) differences for halved time step, BDF2, and tighter nonlinear tolerances, and (d) combined numerical-sensitivity indicators. The dashed 10\% line is only a visual guide, not a statistical confidence or pass/fail threshold. These diagnostics do not redefine the physical peak quantities taken from Fig.~\ref{fig:luminosity}.}
\label{fig:numerical_sensitivity}
\end{figure*}

Fig~\ref{fig:numerical_sensitivity}(a,b) shows that spatial resolution mainly affects the earliest stage, when the heated region is still narrow. The $90\times30$ grid overestimates the sub-day luminosity and the $360\times120$ grid underestimates it relative to the $180\times60$ baseline, but both differences decrease rapidly. By roughly 10 days the three luminosity curves nearly coincide, and thereafter the grid dependence is at the level of a few percent or less.

Only the relative behavior among the numerical variants is used in constructing the sensitivity budget. In particular, Fig.~\ref{fig:numerical_sensitivity} is not used to re-estimate the physical peak time reported from the production multi-EOS light curves in Fig.~\ref{fig:luminosity}.

The time-integration and nonlinear tests in
Fig.~\ref{fig:numerical_sensitivity}(c) are less restrictive than the early coarse-grid comparison. Halving the time step changes the luminosity at approximately the percent level or below through most of the evolution, and tightening the nonlinear tolerances has a negligible effect on the plotted light curve. BDF2 gives the largest difference among these solver variants, but the deviation remains at only a few percent over the displayed interval and falls to well below 1\% by 1000 days. The revised validation therefore does not show the previously inferred large late-time BDF2 drift; the remaining integrator dependence is nevertheless retained in the combined numerical-sensitivity indicator.

The combined numerical-sensitivity indicators in Fig.~\ref{fig:numerical_sensitivity}(d) are $0.61\%$ for peak $T_{\rm s,max}$, $1.53\%$ for the cumulative excess surface-photon energy $E_{\rm surf,ex}$, $1.20\%$ for $L_{\gamma,\rm ex}^{\infty}(1000\,{\rm d})$, $5.33\%$ for the peak excess photon luminosity, and $6.84\%$ for the cumulative excess neutrino energy $E_{\nu,\rm ex}$. With these updated values, the $14.8\%$ BSk26--BSk24 peak-luminosity enhancement is clearly larger than the corresponding $5.33\%$ reference sensitivity scale, as is the BSk22 peak-luminosity deficit, whereas the smaller BSk25 peak contrast remains within that scale. The late-time BSk luminosity differences are more strongly separated from the $1.20\%$ 1000-day sensitivity indicator. These comparisons are used only as a numerical-resolution convention and do not convert the deterministic sensitivity tests into statistical confidence intervals.

With the numerical-sensitivity budget established, we now summarize the main findings and limitations of the controlled EOS comparison.

\FloatBarrier
\section{Conclusions and outlook}
\label{sec:outlook}

The main physical implication is that EOS changes modify the thermal-relaxation pathway rather than simply rescale an outburst light curve. The crossing of the luminosity curves reflects an evolving balance among outward diffusion, inward transport, and crustal storage. The updated BSk factorization shows a predominantly structural control of the early peak surface response, but the late-time luminosity and cumulative neutrino energy exhibit EOS-dependent competition between structural and microphysical contributions rather than a universal single-channel dominance. The similar surface-photon energy fractions but widely different inner-boundary/storage partitions show that internal routing of deposited heat can be more EOS sensitive than the integrated surface output. Within the revised numerical-sensitivity budget, the late-time light-curve differences and the largest peak contrasts are clearly resolved relative to the BSk24 reference scale, while the smallest early peak contrast remains less secure. For observational applications, the late-time light-curve shape therefore remains a particularly robust target in this controlled comparison.

Several simplifications limit the direct applicability and precision interpretation of these results. First, the magnetic field is prescribed, static, and purely poloidal; realistic magnetar fields can contain strong toroidal and multipolar components whose evolution changes the heat-transport pattern. Second, the phenomenological heating prescription does not model the detailed dynamics of crustal failure or magnetospheric return currents. Moreover, fixing the same fractional depth $\xi_h$ does not fix the same density, pressure, composition, or column depth among EOS models. The present results are therefore specific to a fixed-fractional-depth heating experiment; a complementary fixed-density or fixed-column-depth comparison is required to quantify source-location systematics.

Third, the core is represented by a fixed $10^8$~K \emph{local proper} Dirichlet temperature rather than evolved self-consistently with the crust. Since $e^{\Phi(R_{\rm in})}$ is EOS dependent, this boundary condition does not impose a common redshifted core-boundary temperature $\widetilde{T}_{\rm core}$. The large EOS spread in inward energy transfer through the computational inner boundary should therefore be interpreted as a diagnostic of the present fixed-boundary model, not as a fully self-consistent prediction for core heating.

Fourth, the interior thermal diffusion operator and energy quadrature do not include the full general-relativistic lapse, radial-metric, and proper-volume factors. The calculation is partially relativistic: the TOV background, density--radius mapping, surface gravity, and surface luminosity redshift are included, while the interior transport equation retains its Newtonian-form spherical divergence. Because compactness differs among EOS models, the omitted relativistic transport correction is itself EOS dependent and should be quantified before using small percentage-level contrasts as precision EOS discriminants.

Fifth, the evolved thermal domain terminates at $R_{\rm in}$, which is set by the common support of the structure EOS, microphysics EOS, and transport tables and need not coincide with the native crust--core transition $R_{\rm cc}$. Consequently, native crust thickness and computational thermal-domain thickness are not interchangeable. Finally, the smallest early peak-luminosity contrast and some individual factorization components remain comparable to the adopted BSk24 reference sensitivity scale, even though the revised BDF2 comparison stays at only a few-percent level and becomes sub-percent at late times. Evolving magnetic geometry, self-consistent core--crust coupling, fully relativistic transport, alternative heating-depth controls, and source-specific observation operators are therefore natural priorities for future work.

\data{The simulation outputs, channel-resolved neutrino data, and local-frame energy-ledger data that support the findings of this study are available from the corresponding authors upon reasonable request.}

\section*{Software and reproducibility statement}
Each run records the EOS pair, transport table, grid, field geometry, heating parameters, integrator, and numerical tolerances in machine-readable metadata. Because table support sets $R_{\rm in}$, the EOS/table pairing and sampled density range are part of the model definition and should accompany released simulation products.

\section*{Declaration of competing interests}
The authors declare no competing interests.

\bibliographystyle{unsrtnat}
\bibliography{CPC_multiEOS_refs_all_cited}

\end{document}